\documentclass[twocolumn,prc,amssymb, aps,superscriptaddress, 
amsmath]{revtex4}

\usepackage{color}
\usepackage{amsmath,mathtools,booktabs,
            microtype} %,siunitx}

\usepackage{graphicx}
\usepackage{dcolumn}
\usepackage{bm}
\usepackage{multirow}
\usepackage{times}
\usepackage{url}
\usepackage{tikz}
\usepackage[version=4]{mhchem}

\usetikzlibrary{arrows,decorations.pathmorphing}
\usetikzlibrary{arrows.meta}
\usetikzlibrary{calc}

\definecolor{violet}{HTML}{602969}
\definecolor{red}{HTML}{FC0009}
\definecolor{orange}{HTML}{FF6319}
\definecolor{green}{HTML}{00933C}
\definecolor{blue}{HTML}{0036A6}
\definecolor{yellow}{HTML}{FFBE00}
\definecolor{lightgrey}{HTML}{A7A9AC}

\newcommand{\be}{\begin{equation}}
\newcommand{\ee}{  \end{equation}}
\newcommand{\ba}{\begin{eqnarray}}
\newcommand{\ea}{  \end{eqnarray}}

\begin{document}

\title{Statistical Theory of Neutron-Induced Nuclear Fission and
  of Heavy-Ion Fusion}

\author{Hans A. \surname{Weidenm\"uller}}
\email{haw@mpi-hd.mpg.de}
\affiliation{Max-Planck-Institut f\"ur Kernphysik, Saupfercheckweg 1,
  D-69117 Heidelberg, Germany}

\date{\today}

%%%%%%%%%%%%%%%%%%%%%%%%%%%%%%%%%%%%%%%%%555

\begin{abstract}For both reactions I use an approach similar to that
  of compound-nucleus reaction theory. For neutron-induced fission,
  the compound system generated by absorption of the neutron, and the
  nuclear system near the scission point are described as two
  statistically independent systems governed by random-matrix
  theory. The systems are connected either by a barrier penetration
  factor or by a set of transition states above the barrier. Each
  system is coupled to a different set of channels. An analogous model
  is used for heavy-ion fusion. Under the assumption that (seen from
  the entrance channel) the system on the other side of the barrier is
  in the regime of strongly overlapping resonances, closed-form
  analytical expressions for the total probability for fission and for
  fusion are obtained for each value of spin and parity. Parts of
  these expressions can be calculated reliably within existing
  compound-nucleus reaction theory. The remaining parts are the
  probabilities for passage through or over the barrier. These may be
  determined theoretically from the liquid-drop model or
  experimentally from total fission or fusion cross sections.

\end{abstract}
%%%%%%%%%%%%%%%%%%%%%%%%%%%%%%%%%%%%%%%%%%%%%%

\maketitle

%%%%%%%%%%%%%%%%%%%%%%%%%%%%%%%%%%%%%%%%%%%%%%%%%%%%%%%%%%%%%%%%%%%%%%%%

\section{Introduction}
\label{int}

The standard approach to neutron-induced nuclear fission is based on
the liquid-drop model originally proposed by Bohr and
Wheeler~\cite{Boh39}. The capture of a thermal neutron leads to
excitation energies of the compound nucleus close to or above the top
of the fission barrier. The liquid drop deforms and eventually
fissions. The deformation is described in terms of one or several
collective variables. These must be able to pass the fission
barrier. To that end, all or a good fraction of the excitation energy
of the compound nucleus must be transferred to the collective
variables. The barrier is passed via tunneling or via one or several
transition states located on top of the barrier. On the other side of
the fission barrier, the process is reversed, energy of collective
motion is transferred back to the non-collective nuclear degrees of
freedom. As it approaches the scission point, the nuclear system is
again in a highly excited state as witnessed by neutron emission prior
to fission. Most recently that time-dependent picture of the fission
process was formulated theoretically in terms of the
generator-coordinate method in Ref.~\cite{Ber22} where numerous
further references may be found. Extensive numerical work~\cite{Cha11}
shows good agreement with data.

Here I propose a radically different approach. In line with the
standard approach to compound-nucleus reactions, I propose a
statistical model for nuclear fission induced by neutrons, and for
fusion of two heavy ions. The model makes use of the physical picture
described in the previous paragraph. The fissioning (fusioning) system
is separated into two subsytems located on either side of the fission
(fusion) barrier. The two subsystems are connected via barrier
penetration or via a small number of transition states right above the
barrier. The central assumption of the approach is that the two
subsystems can be described as statistically independent members of
the GOE, the time-reversal-invariant Gaussian orthogonal ensemble of
random matrices~\cite{Meh06}. That assumption makes it possible to
derive explicit expressions for total average reaction cross
sections. In detail the model looks as follows.

For a neutron-induced reaction, the incident neutron populates the
compound nucleus (subsystem~1). Barrier penetration or transmission
over the barrier via a small number of transition states connects
subsystem~1 with the nuclear system near the scission point
(subsystem~2). Subsystem~2 undergoes fission, possibly preceded by
neutron emission. Conversely, for the fusion reaction of two heavy
ions, subsystem~2 is the nuclear system reached by the merger of the
two ions. Barrier penetration or transmission over the barrier via a
small number of transition states links subsystem~2 with the compound
nucleus (subsystem~1). The compound nucleus decays by neutron
emission.

Following that picture, the model comprises the following
elements. (i) A set of channels linked to subsystem~1 (the compound
nucleus). For a neutron-induced reaction, that set comprises the
neutron channel. The channels in that set are labeled $(1 a, 2 a,
\ldots)$. (ii) A set of channels linked to subsystem~2 (the fission
channels in case of neutron-induced fission or the channels containing
pairs of heavy ions in the case of the merger of two heavy ions). The
channels in that set are labeled $(1 b, 2 b, \ldots)$. (iii) The
Hamiltonian of the intermediate system. That Hamiltonian consists of
three pieces. The Hamiltonian matrix $H_1$ describes the dynamics of
subsystem~1. The Hamiltonian matrix $H_2$ describes the dynamics of
subsystem~2. The Hamiltonian matrix $H_{\rm tr}$ describes barrier
penetration or transmission over the barrier via a small number of
transition states and links $H_1$ and $H_2$.

The model is solved, and the reaction cross sections are calculated,
under the following assumptions.

(i) The channels labeled $(1 a, 2 a, \ldots)$ are coupled only to
$H_1$. The channels labeled $(1 b, 2 b, \ldots)$ are coupled only to
$H_2$. The two sets of channels do not have any element in common.

(ii) Seen from the entrance channel, the subsystem on the other side
of the barrier is in the regime of strongly overlapping resonances.

(iii) The Hamiltonian matrices $H_1$ and $H_2$ are both governed by
random-matrix theory, i.e., are statistically independent members of
the GOE, the time-reversal-invariant Gaussian orthogonal ensemble of
random matrices~\cite{Meh06}.

Assumption~(i) is plausible because subsystem~2 is strongly deformed
while subsystem~1 is not. Assumption~(ii) implies that the number of
channels strongly coupled to the subsystem on the other side of the
barrier is sufficiently large to permit an asymptotic expansion in
inverse powers of that number. Only the term of leading order is
retained. That assumption makes it possible to give explicit
expressions for reaction cross sections. Without it, numerical
simulations would be required. Assumption~(iii) confines the model to
situations where subsystem~1 and subsystem~2 can be clearly identified
as separate entities, i.e, for excitation energies that are below or
slightly above the top of the fission barrier. Assumption~(iii)
implies that in the vicinity of the actual energy of the reaction, the
local spectral fluctuation properties of the two subsystems coincide
with those of the GOE. (That same assumption is actually used for the
nuclear Hamiltonian in the statistical theory of compound-nucleus
reactions.) For that assumption to hold, the excitation energy in the
compound nucleus (subsystem~1) and in the scissioning system
(subsystem~2) must be larger than several MeV. With neutron binding
energies in the region of several MeV that condition is always met for
neutron-induced reactions. I consider heavy-ion induced fission for
energies where the condition is likewise met. Then the success of the
statistical theory of compound-nucleus reactions~\cite{Mit10, Kaw15}
supports the hypothesis for subsystem~1. It is natural to extend the
hypothesis to subsystem~2, the nuclear system near the scission point
(or point of merger). The compound nucleus created by absorption of
the incident neutron (subsystem~1) and the nuclear system near the
scission point (subsystem~2) are sufficiently different physical
systems (characterized by very different deformations and excitation
energies) to justify the assumption that prior to any coupling their
spectral fluctuations at energy $E$ are statistically
independent. That implies that averages over $H_1$ and $H_2$ can be
done independently.

The approach is implemented with the help of the standard
time-independent formulation of nuclear reaction
theory~\cite{Mah69}. At fixed energy $E$ and for fixed quantum numbers
$(J, \pi)$, the elements of the scattering matrix $S_{1 a, 2 b}(E)$
connect the incident channel labeled $(1 a)$ with the outgoing channel
$(2 b)$ on the other side of the fission barrier. For a
neutron-induced reaction, channel $(1 a)$ carries the incident neutron
and the target nucleus in its ground state, channels~$(1 a', 1 a''
\ldots$) account for neutrons inelastically scattered on the target
nucleus, while channels $(2 b, 2 b', \ldots$) carry pairs of scission
fragments in their ground or excited states but also neutrons emitted
from the nuclear system before it reaches the scission point. For a
fusion reaction, channel $(2 b)$ carries the two ions in their ground
states while channels $(1 a, 1 a', \ldots$) carry a neutron and the
remaining nucleus in its ground state or any excited state or any
other reaction products. Since $H_1$ and $H_2$ are both members of the
GOE, the matrix element $S_{1 a, 2 b}(E)$ is a stochastic process that
depends upon the statistically independent random Hamiltonians $H_1$
for subsystem~1 and $H_2$ for subsystem~2. For the explicit
calculation of the average transition probability $P_{1 a, 2 b}(E) =
\langle |S_{1 a, 2 b}(E)|^2 \rangle$ (the angular brackets denote the
average over both $H_1$ and $H_2$) I consider the following cases,
distinguished by the excitation energies of subsystem~1 and
subsystem~2. Case~(a): The excitation energy of subsystem~1 is in the
regime of isolated compound-nucleus resonances (that case corresponds
to neutron energies of up to $10$ keV or so); case~(b): the excitation
energy of subsystem~1 is in the regime of strongly overlapping
compound-nucleus resonances (the number of channels strongly coupled
to system~1 is large compared to unity); case~($\alpha$): the
excitation energy of subsystem~2 is in the regime of isolated
resonances; case~($\beta$): the excitation energy of subsystem~2 is in
the regime of strongly overlapping resonances (the number of channels
strongly coupled to system~2 is large compared to unity). Which
combination [(a $\alpha$) or (a $\beta$) or (b $\alpha$) or (b
  $\beta$)] of these cases is realized depends, of course, on the
energy of the reaction and on the binding energies of the nuclei
involved. As shown in Refs.~\cite{Wei22, Wei24} it is possible to
derive a closed-form expression for $P_{a b}(E)$ for three out these
four cases with the exception of the case~(a $\alpha$). In case~(b
$\beta$) the resulting expression is completely explicit. In cases~(a
$\beta$) and (b $\alpha$) the expression involves the threefold
integral familiar from the theory of compound-nucleus
reactions~\cite{Ver85}. Case~(a $\alpha$) can only be treated by
numerical simulation.

Assumption~(ii) formulated on the previous page causes the average
transition probability $\langle | S_{1 a, 2 b}(E) |^2 \rangle$ to
factorize. For a neutron-induced reaction, the first factor describes
the probability to reach subsystem~2 from the neutron channel~$(1 a)$
via subsystem~1 and the barrier. The second factor gives the
probability for subsystem~2 to feed channel~$(2 b)$. Such
factorization is the analogue of the factorization of the average
compound-nucleus cross section for strongly overlapping
resonances~\cite{Ver86} where the first (second) factor describes the
probability of formation (decay) of the compound nucleus,
respectively. Factorization implies that the first of the two said
factors is the same for all final channels~$(2 b)$. It gives the total
probability of the reaction to lead to fission, irrespective of what
happens in detail after subsystem~2 has been reached. (Channel~(2 b)
might be a proper fission channel or a channel for neutron emission
prior to fission). That is shown by performing the sum over all final
channels~$(2 b)$. The sum can be done explicitly and gives unity. For
heavy-ion fusion I proceed analogously to determine the total fusion
probability.

In summary, I calculate the average total fission probability $P_{{\rm
    fission}, 1 a}(E)$ and the average total fusion probability
$P_{{\rm fusion}, 2 b}(E)$ for a system of fixed spin~$J$ and
parity$~\pi$ and under the proviso that the resonances in system~1 or
system~2 or both, strongly overlap.  Technically, the average is over
the distribution of the matrix elements of $H_1$ and of
$H_2$. Physically, the average corresponds to an average over a
spectral domain containing a large number of resonances. Multiplying
$P_{{\rm fission}, 1 a}(E)$ and $P_{{\rm fusion}, 2 b}(E)$ with
appropriate geometrical and kinematical factors and summing over $J$
and $\pi$ one obtains the total average reaction cross section for
neutron-induced fission and for heavy-ion fusion, respectively. That
step is standard.

The model used is a variant of a general statistical approach to
transition-state theory, based upon a suggestion in Refs.~\cite{Hag21,
  Ber21} and worked out in Refs.~\cite{Wei22, Wei24}. To make the
paper reasonably self-contained, the model is defined in
Section~\ref{mod}. The statistical assumptions are formulated in
Section~\ref{stat}. Results derived in Refs.~\cite{Wei22, Wei24}) are
collected, applied, and discussed in the following Sections. The
reader who is not interested in technical aspects is advised to skip
Sections~\ref{mod} and \ref{stat}.

\section{Model}
\label{mod}

I confine myself to states of fixed spin and parity and use the
notation of Ref.~\cite{Wei24} throughout. In the
time-reversal-invariant Hamiltonian $H$, subsystem~1 and subsystem~2
are coupled either by the matrix element for tunneling through the
barrier, or by a set of $k$ transition states right above the
barrier. In matrix form, the tunneling Hamiltonian for the first case
is
\ba
\label{m1}
H = \left( \begin{matrix} H_1 & V \cr
  V & H_2 \cr \end{matrix} \right) \ .
\ea
The time-reversal invariant Hamiltonian for transition over the
barrier is given by
\ba
\label{m2}
H = \left( \begin{matrix} H_1 & V_1 & 0 \cr
  V_1^T & H_{\rm tr} & V_2^T \cr
  0 & V_2 & H_2 \cr \end{matrix} \right) \ .
\ea
In both Eqs.~(\ref{m1}, \ref{m2}), $H_1$ ($H_2$) denotes the real and
symmetric Hamiltonian matrix governing subsystem $1$ (subsystem $2$,
respectively), acting in Hilbert space $1$ (in Hilbert space $2$,
respectively). Both Hilbert spaces have dimension $N$. In
Eq.~(\ref{m1}) the matrix $V$ has rank one and carries the tunneling
matrix element ${\cal V}$. In Eq.~(\ref{m2}) the real and symmetric
Hamiltonian matrix $H_{\rm tr}$ acts in transition space and has
dimension $k$. In what follows, the matrix indices $\mu, \mu'$ denote
states in Hilbert space~1 and run from~1 to $N$. The matrix indices
$\nu, \nu'$ denote states in Hilbert space~2. For the model of
Eq.~(\ref{m1}), $\nu, \nu'$ run from $N + 1$ to $2 N$. For the model
of Eq.~(\ref{m2}), these indices run from $N + k + 1$ to $2 N +
k$. The states in transition space are labeled $m, n$. These indices
run from $N + 1$ to $N + k$.

In its most general form, the rank-one coupling matrix in
Eq.~(\ref{m1}) is written as
\ba
\label{m3}
V_{\mu \nu} = (O_1)_{\mu N} {\cal V} (O_2)_{(N + 1) \nu} \ .
\ea
Here $O_1$ and $O_2$ are two arbitrary $N$-dimensional orthogonal
matrices in space~1 (space~2, respectively). The parameter ${\cal V}$
measures the strength of the tunneling process. In Eq.~(\ref{m2}), the
coupling matrices $V_1$ and $V_2$ are real and have $N$ rows and $k$
columns each. The upper index $T$ denotes the transpose. In their most
general form, $V_1$ and $V_2$ can be written as~\cite{Wei24}
\ba
\label{m4}
(V_1)_{\mu m} &=& \sum_{m'} (O_1)_{\mu m'} {\cal V}_{1, m'} (O_{{\rm tr},
  1})_{m m'} \ , \nonumber \\
(V_2)_{\nu m} &=& \sum_{m'} (O_2)_{\nu m'} {\cal V}_{2, m'} (O_{{\rm tr},
  2})_{m m'} \ .
\ea
Here $O_1$ ($O_2$) are $N$-dimensional orthogonal matrices in space~1
(in space~2, respectively) while $O_{{\rm tr}, 1}$ and $O_{{\rm tr},
  2}$ are orthogonal matrices of dimension $k$ in transition
space. The $2 k$ real parameters ${\cal V}_{1, m'}$ and ${\cal V}_{2,
  m'}$ determine the strengths of the couplings.
  
The scattering problem mentioned in Section~\ref{int} is defined by
coupling the states in space~1 to open channels labeled $(1 a, 1 a', 1
a '', \ldots)$ via real matrix elements $W_{1, a \mu}$, those in
space~2 to open channels labeled $(2 b, 2 b', 2 b'', \ldots)$ via real
matrix elements $W_{2, b \nu}$. The relations
\ba
\label{m5}
\sum_\mu W_{1, a \mu} W_{1, a' \mu} &=& \delta_{a a'} v^2_{1 a} \ , \nonumber \\
\sum_\nu W_{2, b \nu} W_{2, b' \nu} &=& \delta_{b b'} v^2_{2 b}
\ea
rule out direct scattering processes $(1 a) \to (1 a')$ and $(2 b) \to
(2 b')$ without intermediary formation of the compound subsystems~1 or
2, respectively. The factor $v^2_{1, a}$ ($v^2_{2 b}$) measures the
strength of the coupling of subsystem~1 to channel $(1 a)$ (of
subsystem~2 to channel $(2 b)$, respectively). The width matrices
\ba
\label{m6}
(\Gamma_1)_{\mu \mu'} &=& 2 \pi \sum_a W_{1, a \mu} W_{1, a \mu'} \ ,
\nonumber \\  
(\Gamma_2)_{\nu \nu'} &=& 2 \pi \sum_b W_{2, b \nu} W_{2, b \nu'} 
\ea
account for multiple backscattering processes $(1 a) \leftrightarrow$
subsystem~1 and $(2 b) \leftrightarrow$ subsystem~2. These matrices
are part of the total width matrix $\Gamma$. For Eq.~(\ref{m1}),
$\Gamma$ is defined by
\ba
\label{m7}
\Gamma = \left( \begin{matrix} \Gamma_1 & 0 \cr
  0 & \Gamma_2 \cr \end{matrix} \right) \ ,
\ea
while for Eq.~(\ref{m2}), $\Gamma$ is given by
\ba
\label{m8}
\Gamma = \left( \begin{matrix} \Gamma_1 & 0 & 0 \cr
  0 & 0 & 0 \cr
  0 & 0 & \Gamma_2 \cr \end{matrix} \right) \ .
\ea
With these definitions, the elements $S_{1 a, 2 b}(E)$ of the
scattering matrix for the reaction $(1 a) \to (2 b)$ at energy $E$
are given by~\cite{Mah69}
\ba
\label{m9}
S_{1 a, 2 b}(E) = - 2 i \pi \sum_{\mu \nu} W_{1, a \mu} D^{-1 }_{\mu \nu}(E)
W_{2, b \nu} \ .
\ea
Here $D^{- 1}(E)$ is the propagator matrix with inverse
\ba
\label{m10}
D(E) = E {\bf 1} - H + (i / 2) \Gamma \ ,
\ea
and ${\bf 1}$ is the unit matrix in the Hilbert space of the
Hamiltonian $H$ defined in Eqs.~(\ref{m1}) or (\ref{m2}).

\section{Statistical Assumptions}
\label{stat}

The elements~(\ref{m9}) of the scattering matrix, although well
defined, cannot be worked out analytically without further simplifying
assumptions on the Hamiltonian $H$. I assume that the excitation
energies of subsystem~1 and of subsystem~2 are both sufficiently large
to justify a statistical treatment. I accordingly assume that the
matrices $H_1$ and $H_2$ are statistically independent members of the
Gaussian Orthogonal Ensemble (GOE) of random matrices. The elements
are zero-centered real Gaussian random variables with second moments
\ba
\label{s1}
\langle (H_1)_{\mu_1 \mu'_1} (H_1)_{\mu_2 \mu'_2} \rangle &=& \frac{\lambda^2}{N}
(\delta_{\mu_1 \mu_2} \delta_{\mu'_1 \mu'_2} + \delta_{\mu_1 \mu'_2} \delta_{\mu'_1
  \mu_2}) \ , \nonumber \\
\langle (H_2)_{\nu_1 \nu'_1} (H_2)_{\nu_2 \nu'_2} \rangle &=& \frac{\lambda^2}{N}
(\delta_{\nu_1 \nu_2} \delta_{\nu'_1 \nu'_2} + \delta_{\nu_1 \nu'_2} \delta_{\nu'_1
  \nu_2}) \ .
\ea
The angular brackets denote the ensemble average. The parameter
$\lambda$ defines the ranges of the two spectra. Eventually, the limit
$N \to \infty$ is taken, with $\lambda$ and $k$ held fixed. The $k$
eigenvalues $E_m$ of $H_{\rm tr}$ and the total energy $E$ of the
system are assumed to be all located near the centers of the spectra
of $H_1$ and $H_2$.

In Refs.~\cite{Wei22, Wei24} it is shown that ${\cal V}$ in
Eq.~(\ref{m3}) must be small in magnitude compared to $\lambda$, and
that the same must hold for all values of $m'$ of the parameters
${\cal V}_{1, m'}$ and ${\cal V}_{2, m'}$ in Eqs.~(\ref{m4}). These
bounds imply that the coupling matrix $V$ in Eq.~(\ref{m1}) and the
coupling matrices $V_1$ and $V_2$ in Eq.~(\ref{m2}) connect mostly
those eigenstates of $H_1$ and of $H_2$ to $H_{\rm tr}$ that lie
within or close to the spectrum of $H_{\rm tr}$. It is only in that
range of energies (which is assumed to be small compared to $\lambda$)
that the statistics of eigenvalues and eigenfunctions of $H_1$ and of
$H_2$ defined via Eqs.~(\ref{s1}) are actually needed. In other words,
Eqs.~(\ref{s1}) quantify the {\it local} spectral fluctuation
properties of $H_1$ and $H_2$ only and do not give a quantitative
parametrization of their overall spectra. That is in line with the use
of random-matrix theory in the statistical theory of nuclear
reactions~\cite{Mit10}.

The average transition probability discussed in Section~\ref{man}
below is parametrized in terms of transmission coefficients $T_{1, a}$
and $T_{2, b}$. These auxiliary parameters are familiar from the
theory of compound-nucleus reactions~\cite{Ver85}. They are defined
for two scattering problems obtained by putting ${\cal V} = 0$ in
Eq.~(\ref{m1}) (or, equivalently, by putting $V_1 = 0 = V_2$ in
Eq.~(\ref{m2})), i.e., for the case without any coupling of
subsystem~1 and subsystem~2. The associated matrices for
backscattering from channel $(1 a)$ to channel $(1 a')$ and for
backscattering from channel $(2 b)$ to channel $(2 b')$ are
\ba
\label{s2}
(S_1)_{1 a, 1 a'}(E) &=& \delta_{a a'} - 2 i \pi \sum_{\mu \mu'} W_{1, a \mu}
      {D^{-1}_1}_{\mu \mu'} W_{1, a' \mu'} \ , \nonumber \\
(S_2)_{2 b, 2 b'}(E) &=& \delta_{b b'} - 2 i \pi \sum_{\nu \nu'} W_{2, b \nu}
      {D^{-1}_2}_{\nu \nu'} W_{2, b' \nu'} \ , \nonumber \\
\ea
where
\ba
\label{s3}
(D_1)_{\mu \mu'}(E) &=& E \delta_{\mu \mu'} - (H_1)_{\mu \mu'} + (i/2)
(\Gamma_1)_{\mu \mu'} \ , \nonumber \\
(D_2)_{\nu \nu'}(E) &=& E \delta_{\nu \nu'} - (H_2)_{\nu \nu'} + (i/2)
(\Gamma_2)_{\nu \nu'} \ ,
\ea
and where $H_1$ and $H_2$ both are members of the GOE and obey the
statistical assumptions~(\ref{s1}). The matrix $S_1(E)$ ($S_2(E)$) is
equal to the scattering matrix for compound-nucleus
scattering~\cite{Ver85} where the compound nucleus is described by the
GOE Hamiltonian $H_1$ ($H_2$, respectively). The transmission
coefficients $T_{1, a}$ and $T_{2, b}$ are defined by
\ba
\label{s4}
T_{1, a} &=& 1 - | \langle (S_1)_{1 a, 1 a}(E) \rangle |^2 \ ,
\nonumber \\
T_{2, b} &=& 1 - | \langle (S_2)_{2 b, 2 b}(E) \rangle |^2 \ .
\ea
The angular brackets denote the average over $H_1$ (over $H_2$,
respectively). These coefficients measure the absorption of flux from
the channel to the compound system or, equivalently, the emission of
flux from the compound system to the channel in question.

The case of strongly overlapping resonances mentioned in
Section~\ref{int} is formally defined for subsystem~1 (for
subsystem~2) by the inequality $\sum_a T_{1, a} \gg 1$ ($\sum_b T_{2,
  b} \gg 1$, respectively). If either of these conditions holds, the
results given below are obtained in terms of an asymptotic expansion
in inverse powers of $\sum_a T_{1, a}$ ($\sum_b T_{2, b}$,
respectively) where only terms of leading order are kept. For the
scattering matrices $S_1(E)$ and $S_2(E)$ that approximation
gives~\cite{Ver86} for $a \neq a'$ and for $b \neq b'$
\ba
\label{s5}
\langle | (S_1)_{1 a, 1 a'}(E) |^2 \rangle &=& T_{1, a} {\cal T}_{1, a'}
        \ , \nonumber \\
\langle | (S_2)_{2 b, 2 b'}(E) |^2 \rangle &=& T_{2, b} {\cal T}_{2, b'} \ .
\ea
The factors
\ba
\label{s6}
      {\cal T}_{1, a'} = \frac{T_{1, a'}}{\sum_{a''} T_{1, a''}} \ , \
      {\cal T}_{2, b'} = \frac{T_{2, b'}}{\sum_{a''} T_{2, b''}}
\ea
give the relative probability for decay of the compound nucleus into
channel $(1 a')$ (channel $(2 b')$, respectively).

\section{Total Transition Probability for Fission and Fusion}
\label{man}

With $S_{1 a, 2 b}(E)$ given by Eq.~(\ref{m9}), the average
transition probability $P_{1 a, 2 b}(E)$ at energy $E$ is given by
\ba
\label{a1}
P_{1 a, 2 b}(E) = \big\langle | S_{1 a, 2 b}(E) |^2 \big\rangle \ .
\ea
The average is over the matrix elements of $H_1$ and of $H_2$. The
symmetry of $S_{1 a, 2 b}(E)$ implies $P_{1 a, 2 b}(E) = P_{2 b, 1 a}(E)$.

It may be instructive to mention why it is not possible to perform the
average in Eq.~(\ref{a1}) analytically without additional simplifying
assumptions. The simpler problem of calculating $\langle | S_{1 a, 1
  a'}(E) |^2 \rangle$ in Eq.~(\ref{s5}) analytically without resorting
to the approximation $\sum_{a'} T_{1, a'} \gg 1$ was
solved~\cite{Ver85} with the help of the supersymmetry
technique. Inspection of Ref.~\cite{Ver85} shows that for two coupled
GOE's, the number of variables in the supersymmetry approach is so
large as to defy an analytical solution. Therefore, the explicit
expressions for the transition probability through or over a barrier
derived in Refs.~\cite{Wei22, Wei24} are obtained with the help of
further simplifying assumptions.

As mentioned in the Introduction I focus attention here on cases
where at least in one of the two subsystems the compound resonances
overlap strongly. The technical aspects of that assumption are defined
at the end of Section~\ref{stat}. Even within that assumption, the
resulting expressions cannot be derived without use of the
supersymmetry technique~\cite{Wei22}. Results derived on that basis in
Refs.~\cite{Wei22, Wei24} are used in Sections~\ref{neu} and
\ref{fus}.

I recall that the total average probability for neutron-induced
fission and that for heavy-ion fusion are given, respectively, by
\ba
\label{a2}
P_{{\rm fission}, 1 a}(E) &=& \sum_b P_{1 a, 2 b}(E) \nonumber \\
P_{{\rm fusion}, 2 b}(E) &=& \sum_a P_{1 a, 2 b}(E) \ .
\ea

\section{Neutron-induced Fission}
\label{neu}

Depending on the energy of the incident neutron I distinguish the
cases where the excitation energy of the compound system (subsystem~1)
is far below, below but close to, or above the height of the fission
barrier. In each of these cases I further distinguish the case where
the resonances in the compound system (subsystem~1) do not overlap or
overlap weakly and the case where they overlap strongly. In all cases
the excitation energy of subsystem~2 is assumed to be in the regime
of strongly overlapping resonances.

\subsection{Excitation energy far below the height of the
  fission barrier}

The tunneling matrix element ${\cal V}$ is so small in magnitude that
lowest-order perturbation theory in ${\cal V}$ suffices. I use
Eq.~(34) of Ref.~\cite{Wei22} and the fact that for strongly
overlapping resonances in subsystem~2 the last factor in that equation
is equal to ${\cal T}_{2, b}$ defined in Eq.~(\ref{s6}). With $\sum_b
{\cal T}_{2, b} = 1$ the first of Eqs.~(\ref{a2}) gives
\ba
\label{n1}
&& P_{{\rm fission}, 1 a}(E) = ({\cal V} / \lambda)^2 \\ 
&& \times \big\langle \big| \sqrt{2 \pi \lambda} \sum_\mu W_{1, a \mu}
[(E - H_1 + (i / 2) \Gamma_1)^{-1 }]_{\mu N} \big|^2 \big\rangle \ .
\nonumber
\ea
Here and in what follows the dimensionless factor $({\cal V} /
\lambda)^2$ is interpreted as the energy-dependent probability $P_{\rm
  tun}(E)$ for tunneling through the fission barrier and written as
\ba
\label{n2}
({\cal V} / \lambda)^2 = P_{\rm tun}(E) \ .
\ea
Then
\ba
\label{n3}
&& P_{{\rm fission}, 1 a}(E) = P_{\rm tun}(E) \\ 
&& \times \big\langle \big| \sqrt{2 \pi \lambda} \sum_\mu W_{1, a \mu}
[(E - H_1 + (i / 2) \Gamma_1)^{-1 }]_{\mu N} \big|^2 \big\rangle \ .
\nonumber
\ea
The factor in angular brackets in Eq.~(\ref{n3}) is given in Eq.~(A4)
of Ref.~\cite{Wei22}. That expression involves the threefold integral
well known from the theory of compound-nucleus
reactions~\cite{Ver85}. For the case of overlapping resonances in
subsystem~1, that factor is equal to ${\cal T}_{1, a}$, and
Eq.~(\ref{n3}) becomes
\ba
\label{n4}
P_{{\rm fission}, 1 a}(E) = {\cal T}_{1, a} P_{\rm tun}(E) \ .
\ea
For first-order perturbation theory to apply, i.e., for Eq.~(\ref{n3})
to hold, $P_{\rm tun}(E)$ must be less than $0.1$ or so because the
correction term is of order $1 / (1 + P_{\rm tun}(E))^2$, see
Eq.~(\ref{n5}).

\subsection{Excitation energy close to but below the height of the
  fission barrier}

If $P_{\rm tun}(E)$ in Eq.~(\ref{n2}) obeys $0.1 \leq P_{\rm tun}(E) <
1$ an analytical solution for $P_{{\rm fission}, 1 a}(E)$ is available
only if the resonances in the compound system (subsystem~1) strongly
overlap. Then Eqs.~(31) and (32) of Ref.~\cite{Wei22} give
\ba
\label{n5}
P_{{\rm fission}, 1 a}(E) = {\cal T}_{1, a} \frac{P_{\rm tun}(E)}{(1 +
  P_{\rm tun}(E))^2} \ .
\ea
The term in the denominator accounts for repeated tunneling back and
forth through the barrier.

\subsection{Passage over the barrier via a set of transition states}

Again, an analytical expression is available only if the resonances in
the compound system (subsystem~1) strongly overlap. Then, Eqs.~(25),
(30), and (16) of Ref.~\cite{Wei24} give
\ba
\label{n6}
P_{{\rm fission}, 1 a}(E) = {\cal T}_{1, a} Y
\ea
where
\ba
\label{n7}
Y = (1 / \lambda^2) \sum_{\mu \nu} \bigg| \sum_{m n} (V_1)_{\mu m}
(G_{\rm tr})_{m n} (V_2)_{\nu n} \bigg|^2
\ea
and
\ba
\label{n8}
G_{\rm tr} &=& ({\bf E} - H_{\rm tr} + i V_1^T V_1 / \lambda + i V_2^T
V_2 / \lambda )^{- 1} \nonumber \\
&=& ({\bf E} - H_{\rm eff})^{- 1} \ .
\ea
Here ${\bf E}$ is the product of the energy $E$ of the reaction and of
the unit matrix in transition space. The operator $G_{\rm tr}$ is the
effective propagator in transition space, and $H_{\rm eff}$ is the
effective Hamiltonian in that space. It consists of the real symmetric
Hamiltonian $H_{\rm tr}$ in Eq.~(\ref{m2}) and an imaginary symmetric
matrix that is due to the coupling of transition space with space~1
and space~2 via the matrices $V_1$ and $V_2$ in Eq.~(\ref{m2}). For a
physical interpretation of $Y$ I introduce the average GOE level
spacing $d = 2 \pi \lambda / N$ and obtain
\ba
\label{n9}
Y = \frac{1}{N^2} \sum_{\mu \nu} \bigg| \frac{2 \pi}{d} \sum_{m n}
(V_1)_{\mu m} (G_{\rm tr})_{m n} (V_2)_{\nu n} \bigg|^2 \ .
\ea
The expression within absolute signs is the dimensionless amplitude
for passing via the transition region from state~$\mu$ in subsystem~1
to state~$\nu$ in subsystem~2, and $Y$ is the average (taken over
states in space~1 and space~2) of the square of that amplitude. In
applications, $d^2$ must be replaced by the product $d_1 d_2$ of the
mean level spacings of subsystem~1 and subsystem~2.

The propagator $G_{\rm tr}(E)$ in Eq.~(\ref{n8}) can be written in
terms of the complex eigenvalues ${\cal E}_l$ with $\Im({\cal E}_l) <
0$ for all $l = 1, \ldots, k$ and the complex orthonormal
eigenfunctions $\psi_l$, $l = 1, \ldots, k$, of $H_{\rm eff}$. The
factor $Y$ in Eq.~(\ref{n7}) takes the form~\cite{Wei24}
\ba
\label{n10}
Y = \frac{1}{N^2} \sum_{m n} \bigg| \sum_l \zeta_{1, m l} \frac{1}{E -
  {\cal E}_l} \zeta_{2, n l} \bigg|^2 \ .
\ea
Except for a factor $(2 \pi / d)^{1/2}$, the parameters $\zeta_{1, m
  l}$ and $\zeta_{2, n l}$ are projections of the eigenfunctions
$\psi_l$ onto the matrices $V_1$ and $V_2$, respectively. The
parameters $\zeta_{1, m l}$ and $\zeta_{2, n l}$ have dimension
(energy)$^{1/2}$. The amplitude within absolute square signs in
Eq.~(\ref{n10}) is the sum of $k$ overlapping Breit-Wigner
resonances. A similar but less general result was obtained in
Ref.~\cite{Hag23} under restrictive assumptions on the matrices $V_1$
and $V_2$.

\subsection{Discussion}
\label{neud}

In deriving expressions~(\ref{n3}) to (\ref{n9}) I have used that
$P_{1 a, 2 b}(E)$ in Eq.~(\ref{a1}) factorizes, one factor being given
by ${\cal T}_{2, b}$. Such factorization is the result of the
statistical assumption~(\ref{s1}) for $H_2$ and of the inequality
$\sum_b T_{2, b} \gg 1$. Factorization makes it possible to sum over
exit channels explicitly and to obtain closed expressions for $P_{{\rm
    fission}, 1 a}(E)$. Eqs.~(\ref{n4}, \ref{n5}) and (\ref{n6}) show
that factorization likewise holds with respect to the entrance-channel
dependence of $P_{{\rm fission}, 1 a}(E)$, owing to the statistical
assumption~(\ref{s1}) on $H_1$ and the inequality $\sum_a T_{1, a} \gg
1$. In all these cases, the orthogonal invariance of the GOE removes
all reference to intermediate states in subsystem~2 and subsystem~1
and leaves us for $P_{{\rm fission}, 1 a}(E)$ with a first factor that
depends only upon transmission coefficients, and a second factor that
depends only upon the dynamics of the transition through or over the
barrier. Factorization in Eq.~(\ref{n3}) is the result of first-order
perturbation theory and the statistical assumption~(\ref{s1}) on $H_1$
without any additional conditions on the resonances in subsystem~1.

When written in the form
\ba
\label{n11}
\langle | (S_1)_{1 a, 1 a'}(E) |^2 \rangle &=& {\cal T}_{1, a} T_{1, a'} \ ,
\ea
Eq.~(\ref{s5}) for the average compound-nucleus reaction probability
displays a striking similarity to $P_{{\rm fission}, 1 a}(E)$ in
Eqs.~(\ref{n4}, \ref{n5}) and (\ref{n6}). The factor ${\cal T}_{1, a}$
appears in all these expressions and accounts for compound-nucleus
formation. The factor $T_{1, a'}$ in Eq.~(\ref{n11}) (which accounts
for the decay of the compound nucleus) is in Eqs.~(\ref{n4}, \ref{n5})
and (\ref{n6}) replaced by a factor that is given in terms of $P_{\rm
  tun}(E)$ or of $Y$ and accounts for transmission through or over the
barrier. The difference between Eq.~(\ref{n11}) and the fission
probability is that in Eq.~(\ref{n11}) the sum in the denominator of
${\cal T}_{1, a}$ extends over all channels. In $P_{{\rm fission}, 1
  a}(E)$ that same sum does not comprise the fission channel. With
that slight proviso the results of the paper show that neutron-induced
fission may be seen as a compound-nucleus reaction that feeds the
fission channel.

For practical applications it is important that the factor ${\cal
  T}_{1, a}$ in Eqs.~(\ref{n4}, \ref{n5}), amd (\ref{n6}) occurs
likewise in the parametrization of the compound-nucleus cross section
for strongly overlapping resonances in Eqs.~(\ref{s5},
\ref{s6}). Calculations of compound-nucleus reaction cross sections
are completely standard. The factor ${\cal T}_{1, a}$ is, therefore,
known very precisely and easily available. The factor in the second
line of Eq.~(\ref{n1}) involves the threefold integral of
compound-nucleus reaction theory. It can be obtained by a slight
modification of the standard numerical program for that
integral. Thus, the channel-dependent factors in Eqs.~(\ref{n1},
\ref{n4}, \ref{n5}) and (\ref{n6}) can all be calculated reliably
within standard compound-nucleus reaction theory.

In principle, the probability $P_{\rm tun}(E)$ in Eqs.~(\ref{n3}) and
(\ref{n4}) can be computed semiclassically with the help of the WKB
approximation and the use of collective variables. For spontaneous
fission, that is the standard procedure, see Ref.~\cite{Sch22} and
references therein. However, as stated in the Introduction, the
present theory deals with states of fixed spin $J$ and parity $\pi$ of
the neutron-induced fission reaction and $P_{\rm tun}(E)$, written
explicitly as $P_{\rm tun}(E; J, \pi )$, must, in principle, be
determined separately for each pair of values $(J, \pi)$. I expect
that collective motion in general and, therefore, $P_{\rm tun}(E; J,
\pi)$ in particular are fairly independent of $(J, \pi)$. I, thus,
expect that $P_{{\rm fission}, 1 a}(E)$ in Eqs.~(\ref{n3}, \ref{n4})
and (\ref{n5}) can be estimated fairly reliably. For the case of
passage over the barrier, the parameters ${\cal E}_l$ and $\zeta_{1, m
  l}, \ \zeta_{2, m l}$ in Eq.~(\ref{n10}) are also determined by
collective features of the system. Therefore, I expect that here, too,
$Y$ is nearly independent of spin and parity of the system. A
prediction of value and energy dependence of $Y$ has to be based upon
a theoretical model for the transition Hamiltonian $H_{\rm tr}$, i.e.,
a collective dynamical model for the transition states and their
coupling to the states in subsystem~1 and subsystem~2. That seems
realistic only for $k = 1$ and $k = 2$ because the number of
parameters in Eq.~(\ref{n10}) increases strongly with increasing $k$.

The total cross section $\sigma_{\rm fission}(E)$ for neutron-induced
fission at energy $E$ is obtained by multiplying the fission
probability $P_{{\rm fission}, 1 a}(E; J, \pi)$ (now written with its
full dependence on $J$ and $\pi$) with an entrance-channel dependent
factor $C_{1, a}(E; J, \pi)$ that depends upon geometrical and
kinematical factors, and summing over both values of $\pi$ and all
values of $J$ participating in the reaction,
\ba
\label{n12}
\sigma_{\rm fission}(E) = \sum_{J, \pi} C_{1, a}(E; J, \pi) P_{{\rm fission},
  1 a}(E; J, \pi) \ .
\ea
Here $C_{1, a}(E; J, \pi)$ is not written in full because that
expression is well known, is lengthy, and only deflects attention from
the essential elements of the theory. The explicit expression may be
found, for instance, in Ref.~\cite{Lan58}.

For tunneling through the barrier an example is provided by
Eq.~(\ref{n5}). For brevity ${\cal P}_{\rm tun}$ is defined as
\ba
\label{n12a}
{\cal P}_{\rm tun}(E; J, \pi) = \frac{P_{\rm tun}(E; J, \pi)}{(1 +
P_{\rm tun}(E; J, \pi)^2} \ .
\ea
Use of Eq.~(\ref{n12a}) in Eq.~(\ref{n5}) and insertion of the latter
into Eq.~(\ref{n12}) gives
\ba
\label{n13}
\sigma_{\rm fission}(E) &=& \sum_{J \pi} C_{1, a}(E; J, \pi) {\cal T}_{1,
  a}(J, \pi) {\cal P}_{\rm tun}(E; J, \pi) \ . \nonumber \\
\ea
If $P_{\rm tun}(E; J, \pi)$ depends upon $(J, \pi)$ only weakly,
\ba
\label{n14}
P_{\rm tun}(E; J, \pi) \approx P_{\rm tun}(E) \ ,
\ea
Eq.~(\ref{n13}) takes the form
\ba
\label{n15}
\sigma_{\rm fission}(E) &\approx& \bigg( \sum_{J \pi} C_{1, a}(E; J, \pi)
{\cal T}_{1, a}(J, \pi) \bigg) \ {\cal P}_{\rm tun}(E) \nonumber \\
&=& \sigma_{\rm compound}(E) \ {\cal P}_{\rm tun}(E) \ .
\ea
The last line defines the total cross section $\sigma_{\rm
  compound}(E)$ for compound-nucleus formation. The value of
$\sigma_{\rm compound}(E)$ is completely determined by nuclear
reaction data and can be calculated reliably. The ratio $\sigma_{\rm
  fission}(E) / \sigma_{\rm compound}(E)$ yields ${\cal P}_{\rm
  tun}(E)$. If the approximation~(\ref{n14}) does not apply, the ratio
\ba
\label{n16}
\sigma_{\rm fission}(E) / \sigma_{\rm compound}(E) = \langle {\cal P}_{\rm
  tun}(E; J, \pi) \rangle
\ea 
gives the weighted average of ${\cal P}_{\rm tun}(E; J, \pi)$ taken
over the values of $(J, \pi)$ that participate in the
reaction. Analogous conclusions hold for the results in
Eqs.~(\ref{n3}), (\ref{n5}), and (\ref{n6}).

The values of ${\cal P}_{\rm tun}(E)$, of $\langle {\cal P}_{\rm
  tun}(E; J, \pi) \rangle$, of $Y$, and of $\langle Y \rangle$
obtained from data on total fission cross sections may serve as tests
for collective models of barrier penetration. That is true, in
particular, of the energy dependence of $P_{\rm tun}(E)$ which relates
directly to the shape of the barrier. For small neutron energies,
$P_{\rm tun}(E)$ should be close to the barrier penetration factor for
spontaneous fission in neighboring nuclei.

\section{Heavy-ion Fusion}
\label{fus}

As an example I consider the case where the excitation energies both
of subsystem~1 and of subsystem~2 are in the regime of overlapping
resonances. Other cases may be discussed in analogy to
Section~\ref{neud}. I use the second of Eqs.~(\ref{a2}). For the case
of barrier penetration, summation of the result in Eq.~(32) of
Ref.~\cite{Wei22} over channels $(1 a)$, the first of Eqs.~(31) of
Ref.~\cite{Wei22}, and Eq.~(\ref{n12a}) give
\ba
\label{f1}
P_{{\rm fusion}, 2 b}(E) = {\cal T}_{2, b} {\cal P}_{\rm tun}(E)
\ea
while for barrier transition via a set of $k$ transition states
Eq.~(25) of Ref.~\cite{Wei24} gives
\ba
\label{f2}
P_{{\rm fusion}, 2 b}(E) = {\cal T}_{2, b} Y \ .
\ea
Here $P_{\rm tun}(E)$ and $Y$ are defined in Eqs.~(\ref{n2}) and
(\ref{n7}), respectively.

For practical applications, numerator and denominator of the factor
${\cal T}_{2, b}$ in Eqs.~(\ref{f1}) and (\ref{f2}) are given for each
value of $J$ and $\pi$ in terms of the transmission coefficients
$T_{2, b'}$ for channels~(2 b') containing two heavy ions each in the
ground or in an excited state, see Eqs.~(\ref{s6}). These transmission
coefficients can be calculated from the optical model for elastic
scattering. The optical-model potential for scattering of two heavy
nuclei seems less well investigated both phenomenologically and
theoretically than that for nucleon-nucleus scattering, see, for
instance, Refs.~\cite{Kho16, Fur19}. Evaluation of the denominator in
${\cal T}_{2, b}$ requires, in addition, the average level densities
of the fission products. These are, in general, known as well as that
of the target nucleus in a neutron-induced reaction. Thus, the factor
${\cal T}_{2, b}$ in Eqs.~(\ref{f1}, \ref{f2}) can perhaps not be
calculated as precisely as ${\cal T}_{1, a}$ for a neutron-induced
fission reaction but can be predicted at least semiquantitatively from
known nuclear data.

As for the factors that determine penetration through or transition
over the barrier, much of the discussion in Section~\ref{neud} applies
in the present case as well and is not repeated here.

The total average fusion probability in Eqs.~(\ref{f1}, \ref{f2}) is
similar in form to the probability of a compound-nucleus reaction that
starts in channel~(2 b) and feeds transmission through or over the
barrier. The probability for that to happen is determined by the
factors $P_{\rm tun}(E)$ and $Y$. In full analogy to
Section~\ref{neud}, these may be determined theoretically with the
help of collective variables.

In analogy to Eq.~(\ref{n12}), the total fusion cross section is
written as
\ba
\label{f3}
\sigma_{\rm fusion}(E) = \sum_{J \pi} C_{2, b}(E; J, \pi) P_{{\rm fusion},
  2 b}(E; J, \pi) \ .
\ea
With the help of Eq.~(\ref{f1}) that gives
\ba
\label{f4}
\sigma_{\rm fusion}(E) = \sum_{J \pi} C_{2, b}(E; J, \pi) {\cal T}_{2,
  b}(J, \pi) {\cal P}_{\rm tun}(E; J, \pi) \ ,
\ea
and correspondingly for Eq.~(\ref{f2}). The cross section for merger,
defined as
\ba
\label{f5}
\sigma_{\rm merger}(E) = \sum_{J \pi} C_{2, b}(E; J, \pi) {\cal T}_{2,
  b}(J, \pi) \ ,
\ea
is available from nuclear data. It follows that ${\cal P}_{\rm
  tun}(E)$ or $\langle {\cal P}_{\rm tun}(E; J, \pi) \rangle$ may be
determined from the ratio of the measured total fusion cross section
and the cross section for merger, and correspondingly for $Y$ and
$\langle Y \rangle$.

\section{Conclusions}

For neutron-induced fission, I have modeled both the compound nucleus
generated by absorption of the incident neutron and the nuclear system
near the scission point as two statistically independent dynamical
systems that can be described in terms of random-matrix theory. For
heavy-ion fusion, I have used the same model for the compound system
reached after merger of the two ions, and for the compound nucleus at
the other side of the barrier. In both cases I have assumed that,
seen from the entrance channel, the system on the other side of the
barrier is in the regime of strongly overlapping resonances. That
makes it possible to derive, for fixed quantum numbers $(J, \pi)$,
explicit expressions for the average probability for neutron-induced
fission and for heavy-ion fusion leading to specific final
channels. These expressions factorize. The first factor describes the
reaction up to the point where the system has passed the barrier. The
second factor describes how the resulting highly excited system
de-excites and/or decays into fragments. Because of the statistical
assumptions these two factors are completely independent of each
other, in the same sense in which the decay of the compound nucleus in
the regime of strongly overlapping resonances is independent of its
mode of formation.

Summation over all final channels yields the total average
probabilities in Eqs.~(\ref{a2}) for neutron-induced fission or
heavy-ion fusion. These expressions are independent of what happens
after the system has passed the barrier. The expressions~(\ref{a2})
relate the probability for neutron-induced fission and for heavy-ion
fusion to elements of the statistical theory of nuclear
reactions. Parts of these expressions can, therefore, be calculated
accurately within existing compound-nucleus theory. The parts of the
theory that go beyond the standard statistical theory are the
probabilities for penetration through or passage over the fission
barrier. These must be calculated within a model for collective motion
or, alternatively, may be determined experimentally from total fission
or fusion cross sections. In that way the present approach promises
direct insight into the dynamics of barrier penetration and of
transmission over the barrier.

The present theory does not specify what happens after the system has
passed the barrier. In principle, that question must be addressed
separately and independently for every value of $(J, \pi)$. It is
conceivable, however, that collective dynamics beyond the barrier
provides a description independent of $(J, \pi)$. In that case, the
fission (fusion) reaction would be described by two factors: The total
cross section for fission (fusion) (given in the present paper) would
be multiplied by the probability (given by the collective dynamics)
for the system on the other side of the barrier to decay into a series
of reaction products.

%%%%%%%%%%%%%%%%%%%%%%%%%%%%%%%%%

\end{document}